\documentclass[]{spie}  

\usepackage{amsmath,amsfonts,amssymb}
\usepackage{graphicx}
\usepackage[colorlinks=true, allcolors=blue]{hyperref}

\title{``Fast" and Furious focal-plane wavefront sensing at W.~M.~Keck~Observatory\thanks{The data presented herein were obtained at the W. M. Keck Observatory, which is operated as a scientific partnership among the California Institute of Technology, the University of California and the National Aeronautics and Space Administration. 
The Observatory was made possible by the generous financial support of the W. M. Keck Foundation.}}

\author[a]{Steven~P.~Bos}
\author[b]{Michael~Bottom}
\author[c]{Sam~Ragland}
\author[c]{Jacques-Robert~Delorme}
\author[c]{Sylvain~Cetre}
\author[d]{Laurent~Pueyo}

\affil[a]{Leiden Observatory, Leiden University, P.O. Box 9513, 2300 RA Leiden, The Netherlands}

\affil[b]{Institute for Astronomy, University of Hawaii, 640 N. Aohoku Place, Hilo, HI 96720, USA.}

\affil[c]{W. M. Keck Observatory, 65-1120 Mamalahoa Highway., Kamuela, HI 96743, USA.}

\affil[d]{Space Telescope Science Institute, Baltimore, MD 21218, USA}

\authorinfo{Send correspondence to Michael Bottom, mbottom@hawaii.edu}

\begin{document} 
\maketitle

\begin{abstract}
High quality, repeatable point-spread functions are important for science cases like direct exoplanet imaging, high-precision astrometry, and high-resolution spectroscopy of exoplanets.
For such demanding applications, the initial on-sky point-spread function delivered by the adaptive optics system can require further optimization to correct unsensed static aberrations and calibration biases.
We investigated using the Fast and Furious focal-plane wavefront sensing algorithm as a potential solution.
This algorithm uses a simple model of the optical system and focal plane information to measure and correct the point-spread function phase, without using defocused images, meaning it can run concurrently with science.  
On-sky testing demonstrated significantly improved PSF quality in only a few iterations, with both narrow and broadband filters.  
These results suggest this algorithm is a useful path forward for creating and maintaining high-quality, repeatable on-sky adaptive optics point-spread functions.  
\end{abstract}

\section{Introduction}\label{sec:intro} 
Adaptive optics (AO) systems have revolutionized ground-based astronomy by improving the angular resolution and power density of astronomical images beyond the limits imposed by atmospheric turbulence.  Current AO systems can operate close to the diffraction limit in terms of point-spread function (PSF) size, meaning the telescope diameter ultimately determines the image sharpness.  At these regimes of very high quality AO correction, unsensed aberrations in the system coming from optics after the wavefront sensor increasingly limit the PSF quality.  Several science cases benefit from further correction beyond this limit.  For example, in high-contrast imaging, these aberrations show up as bright ``speckles'' in the focal plane at angular separations similar to the planets of interest, limiting sensitivity.  Additionally, these kinds of distortions limit the repeatability of the PSF from night to night, which can affect the systematic error floor for science cases like high-precision astrometry.  Finally, a number of science cases such as high precision radial velocities\cite{crepp2016ilocater} and high contrast spectroscopy\cite{mawet2017fiber} depend on coupling telescope light into single-mode fibers, which requires precise control of both the position and electric field of the PSF for acceptable throughput.

These unsensed optical aberrations evolve slowly, unlike atmospheric turbulence, and can thus in principle be corrected by changing the setpoint of the deformable mirror to compensate them.  However, finding the correct setpoint is not straightforward.  Even in the simplest cases, such as a PSF slightly out of focus, it is not clear whether to add or subtract a focus term; this is the well-known ``sign ambiguity'' in optics.  Additionally, different aberrations can be present at once and may be difficult to disentangle individually.

The simplest method to remove these quasi-static errors is to pick a basis to describe the PSF (such as the Zernike modes), then step through each mode and try to adjust the amount of power in the mode to see whether the PSF improves or not.  This ``Zernike tuning'' is inefficient and tedious.  More sophisticated strategies use the deformable mirror to generate fixed, known patterns that interfere with the underlying aberrations in the image and thus allow for breaking the sign ambiguity while sensing all the aberrations at once.  The most popular of these are based on the Gerchberg-Saxton algorithm,\cite{fienup1982phase, gerchberg1972practical} which uses two or more images at different focal positions to determine the aberrations in the system.  In astronomy, the Gerchberg-Saxton algorithm is commonly used to tune up the point-spread function \textit{before} opening the telescope, so as to determine a good initial setpoint for the adaptive optics system.  It has been deployed at Palomar\cite{burruss2010demonstration}, LBT \cite{bechter2019characterization}, and other telescopes, and is currently the in-house algorithm at W.M. Keck observatory run before every adaptive optics night.\cite{ragland2016point}

However, in all these cases the algorithm runs off-sky using a synthetic light source rather than starlight, which necessarily means some non-common path optical errors, including from the primary,  secondary, and sometimes tertiary mirrors.  This is particularly true in the case of segmented telescopes like Keck, where piston-type errors are invisible to wavefront sensors that measure slopes of the light wave, like in Shack-Hartmann designs.  As such, while the initial setpoint usually delivers a good PSF, there are still unsensed wavefront errors in the system that prevent the ultimate performance. 

It is possible to run the Gerchberg-Saxton algorithm on-sky, using a star as the light source, but this presents its own set of challenges.  The most basic challenge is intrinsic to the algorithm, which requires images taken at different focal planes, typically in front of and behind the focal position.  The point-spread function is enlarged at these positions, so the signal-to-noise ratio per pixel drops, especially in the presence of sky background light.  Therefore, longer exposure times or frame averaging must be used to increase the signal-to-noise ratio.  Further iterations use precious on-sky time; experiments at Keck typically produce tune-up times of 20-30 minutes using long-exposure phase diversity.\cite{mugnier2008line} The defocused frames are of no scientific utility, so the PSF sharpening procedure is a sunk cost.  Large telescope slews or changes to the science instrument state can require redoing the calibration.

A more ideal approach to on-sky PSF correction would improve on these shortcomings.  First, it would work on-sky without significant degradation in signal-to-noise ratio.  Second, it would not use diversity frames.  Third, it would converge quickly in terms of wall clock time and be robust to non-idealities in the system.

A good candidate to fulfill these needs is the Fast and Furious Wavefront Sensing algorithm\cite{keller2012extremely, korkiakoski2014fast} (aka Fast and Furious, or F\&F).
Fast and Furious works in the loop with the adaptive optics system, and effectively requires no diversity frames, using the \textit{previous} correction applied to break the sign ambiguity.  
It thus does not require any defocusing or corruption of the science images, leading to high efficiency.
It can operate continuously while observing, thus being robust to slow drifts in the optical system.
Finally, it works with both narrow and broadband light, and only requires a simple model of the pupil of the optical system.

The Fast and Furious algorithm was demonstrated on-sky recently on Subaru's SCExAO instrument as a promising method of controlling the ``low-wind effect'' \cite{bos2020sky}, an annoying situation where one or more areas of the pupil located between consecutive spiders have discrete phase piston errors, believed to be due to thermal nonuniformities in the secondary mirror supports.\cite{milli2018low} Adaptive optics systems often have trouble sensing these kinds of errors and thus a focal-plane approach was developed.
In this work, we port the algorithm over to Keck and demonstrate good performance, with rapid and stable convergence using broad and narrowband light over a range of wavelengths, and robust performance with poor atmospheric conditions.  
This is the first demonstration of Fast and Furious with a segmented aperture telescope, and the second demonstration on-sky.  
Our results support the conclusion that Fast and Furious is a promising method for robust on-sky PSF optimization in adaptive optics.

The outline of this paper is as follows: in Section \ref{sec:FF}, we will describe the Fast and Furious algorithm.  In Section \ref{sec:perfmetric}, we describe performance metrics to quantify PSF quality.  In Section \ref{sec:FFKeck}, we will present our particular implementation at Keck, including bench results validating its performance against the Gerchberg-Saxton algorithm.  In Section \ref{sec:results}, we present our on-sky tests.  Finally, we discuss our results with a view towards further work and broader implementation.

\section{Fast \& Furious}\label{sec:FF} 
\begin{figure}[!htb]
\begin{center}
\begin{tabular}{c}
\includegraphics[width=0.75\textwidth]{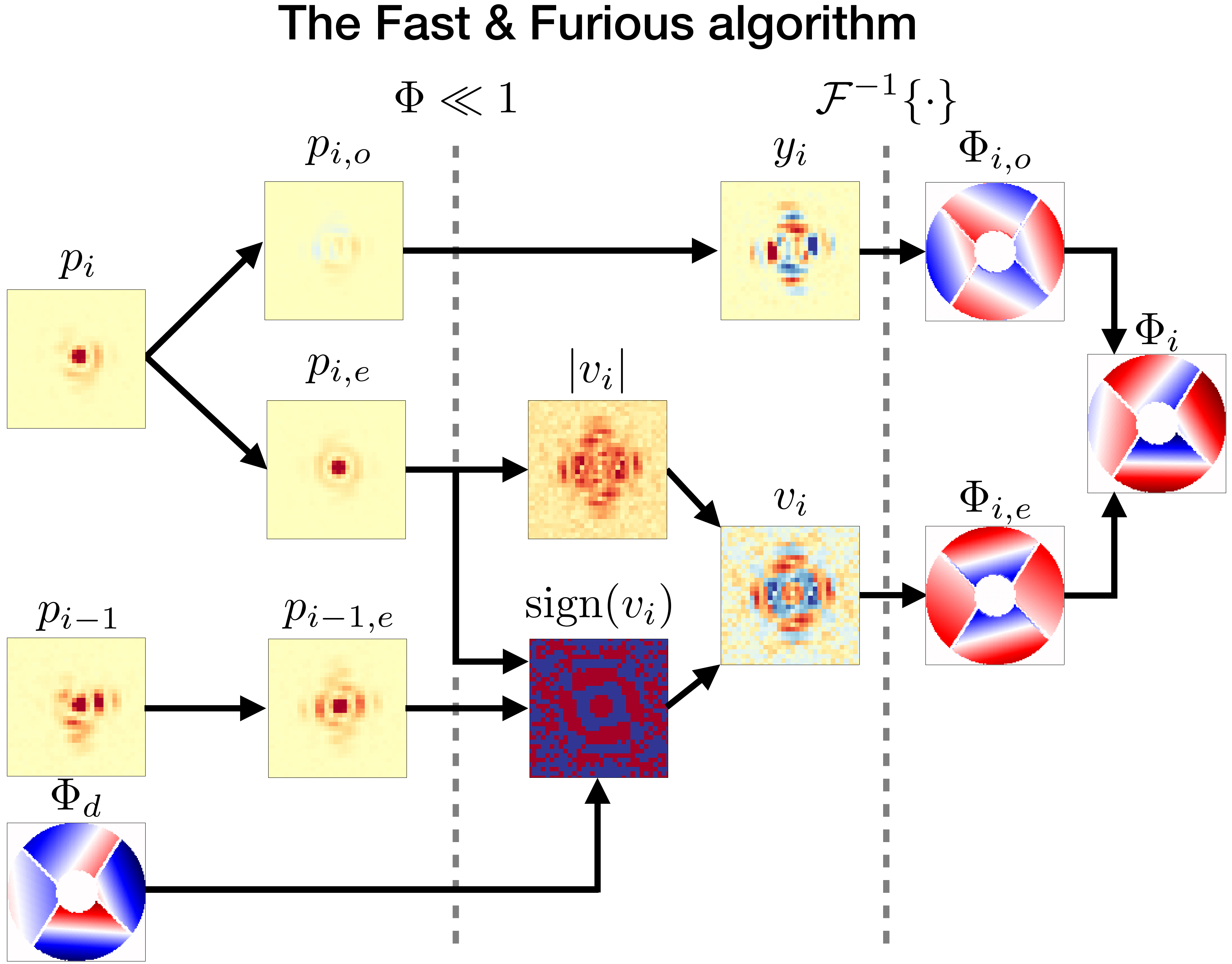}
\end{tabular}
\end{center}
\caption{
Overview of the Fast and Furious focal-plane wavefront sensing algorithm. 
Adopted from Ref~\citenum{bos2020sky}.
}
\label{fig:FF_explanation}  
\end{figure}

Here we give a brief overview of the Fast and Furious focal-plane wavefront sensing algorithm (F\&F); more in-depth details are in references \citenum{keller2012extremely, korkiakoski2014fast, wilby2016fast, wilby2018laboratory}, and \citenum{bos2020sky}.
F\&F is an extension of the sequential phase diversity algorithm introduced by Ref~\citenum{gonsalves2002adaptive}.
Conventional phase diversity algorithms break the sign degeneracy of even phase modes by using in- and out-focus images \cite{gonsalves1982phase, paxman1992joint}.   
Sequential phase diversity algorithms use previous DM commands as diversity, which makes these algorithms more efficient as no observing time is lost to recording defocused images.
F\&F's extension allows it to operate in regimes with higher wavefront aberrations compared to the original sequential phase diversity algorithm. \\
\\ 
\autoref{fig:FF_explanation} presents a graphical overview of the algorithm. 
The equations used by F\&F to derive the phase estimate are based on the following assumptions:  (0) the exit pupil of the instrument is real and symmetric; (1) the PSF is a real ``point''-spread function, not an image of an extended object; (2) the PSF is formed by monochromatic light; (3) the PSF is distorted by phase-only aberrations; (4) these aberrations are small ($\theta \ll 1$ radian); (5) the PSF is shift invariant, which means that the algorithm cannot be combined with focal-plane coronagraphs. 
Some of these assumptions may be relaxed in practice while still maintaining good performance; only the assumption of shift-invariance is mandatory.  
Every iteration of F\&F performs the following steps. 
First, the image ($p_i$) is split into its even ($p_{i,e}$) and odd ($p_{i,o}$) components by using Fourier transform symmetries. 
The image's odd component directly solves for the odd focal-plane electric field ($y_i$), and via an inverse Fourier transform, the odd pupil-plane phase ($\Phi_{i,o}$).
In a similar process, the image's even component is used to derive the absolute value of the even focal-plane electric field ($|v_i|$). 
The sign of the even focal-plane electric field (sign($v_i$)) is solved by using the image's even component, the previous image's even component ($p_{i-1,e}$), and the applied DM command of the previous iteration ($\Phi_d$). 
This then leads to an estimate of the even focal-plane electric field ($v_i$), and the even pupil-plane phase. 
The estimates of the odd and even pupil-plane phase are combined to get the total phase estimate. 
It is possible to project this phase estimate then on a mode basis of choice to target specific aberrations or alleviate the lower signal-to-noise ratio of high-order modes. 
In this work we project onto low-order Zernike modes.
This process is repeated for every iteration.

\section{Performance metrics}\label{sec:perfmetric}
To quantify our the performance of F\&F we adopt two metrics from Ref~\citenum{bos2020sky}: the Strehl ratio approximation ($SRA$) and the variance of the normalized first Airy ring ($VAR$). 
The $SRA$ estimates the Strehl ratio by using a modified encircled energy metric. 
It compares the data $p$ with a numerical PSF $|a|^2$. 
The $SRA$ is calculated as: 
\begin{equation}\label{eq:strehl_measurement}
SRA = \frac{p(r < 1.22 \ \lambda / D)}{p(r < 11.5 \ \lambda / D)} \cdot \frac{|a|^2(r < 11.5 \ \lambda / D)}{|a|^2(r < 1.22 \ \lambda / D)},
\end{equation}
with $\lambda$ the central wavelength of the filter and $D$ the diameter of the telescope. 
The $SRA$ estimates are affected somewhat by the position of the PSF in the image window. 
For example, if the PSF is in the corner of the image the $SRA$ might be slightly different compared to when the same PSF is in the center. 
Therefore, $SRA$ measurements should generally be compared to measurements in the same test. 

The $VAR$ measures the quality of the first Airy ring.
This allows us to measure the effect of low-order aberrations on the PSF. 
This metric can be useful in situations where the AO residuals dominate $SRA$ changes, but that F\&F still corrects low-order aberrations\cite{bos2020sky}.
The $VAR$ is calculated as:
\begin{equation}\label{eq:VAR}
VAR = \text{Var} \biggl( \frac{p(1.52 \ \lambda / D < r < 2.14 \ \lambda / D)}{\langle p(1.52 \lambda / D < r < 2.14 \ \lambda / D) \rangle} \cdot
\frac{\langle |a|^2(1.52 \ \lambda / D < r < 2.14 \ \lambda / D) \rangle}{|a|^2(1.52 \ \lambda / D < r < 2.14 \ \lambda / D)}  \biggr) 
\end{equation}  
An unaberrated PSF will have $VAR=0$, while an aberrated PSF will have $VAR > 0$.  

\section{Implementation at Keck/Nirc2}\label{sec:FFKeck} 

\subsection{System hardware overview}
The Keck II adaptive optics system\cite{wizinowich2000first} consists of a 349-actuator deformable mirror conjugated to the primary mirror, a Shack-Hartmann wavefront sensor, and a real-time controller.  A recent upgrade introduced a pyramid wavefront sensor\cite{bond2020adaptive} which optimizes performance in the infrared for stars such as M dwarfs.  In all results that follow, we used the facility adaptive optics science camera, NIRC2, as our focal-plane sensor.  NIRC2 is an infrared imager operating from 0.9 - 5.3 $\mu$m, with multiple options for grism spectroscopy and coronagraphy.  We set the plate scale to 10 milliarcseconds/pixel, though several other options are possible.

While NIRC2 has competitive performance,\cite{ghez2008measuring,nobelghez} it is not a particularly fast camera, with electronics about 20 years old.  As such, frame times are rarely shorter than 10 seconds for extended fields even when subframing.  (This motivates the entertaining quotes around the word ``Fast'' in the title.)  A related point is that the convergence time of the Fast and Furious algorithm during lab tests has only been 10-20 frames, while previous on-sky results have resulted in convergence times of hundreds of frames.  Therefore, it was unclear whether our implementation would be practically useful: hundreds of frames is unacceptable, while 10-20 frames can be accomplished in less than five minutes.

\subsection{Hardware control software}

The Keck control system is based on the Experimental Physics and Industrial Control System (EPICS) architecture originally developed at Los Alamos National Lab. \cite{dalesio1991epics}  Several python libraries have been written to interact with the underlying Keck real-time controller without requiring low-level control of EPICS channels, in order to simplify writing high-level algorithms, such as F\&F.  In our case, we use one module that interfaces with NIRC2 and one that interfaces with the deformable mirror.  The latter allows for control by sending 21$\times$21 matrices, which are automatically converted into either voltages or wavefront sensor offsets depending on whether the AO loops are closed.  Tip-tilt offsets are also possible inputs, but these were not used for this implementation.  Currently, there are separate modules for running the pyramid sensor and Shack-Hartmann sensor, but this is expected to be unified in the near future.  These libraries were initially developed to control the modules of the Keck Planet Imager and Characterizer\cite{jovanovic2020} and related components. 

\subsection{Loop control software}
The F\&F loop control software uses the same scripts as in Ref~\citenum{bos2020sky}, with some minor modifications.  First, due to the thermal background and detector systematics in NIRC2, an image cleaning subroutine was implemented which allowed for background subtraction, bad pixel removal, and flat correction; optionally, this could be performed automatically by looking at the image statistics without having to take extra images.  Second, some parts of the code were refactored to use human-readable configuration files to set up the system rather than modifying any code directly.  The \textit{configobj} module which interprets these configuration files checks for typos and parameter range errors (eg, typing 1.6 instead of 1.6e-6 for the wavelength) which prevents time on-sky from being wasted due to these avoidable mistakes.

The F\&F algorithm is model-based, that is, it requires some knowledge of the optical system pupil to work.  We generated maps of the different pupil masks in the system using measured values from Keck and NIRC2, including the fixed, medium, and large hexagonal masks, and the open configuration which only uses the primary mirror as the mask.  Similarly, as the algorithm requires a particular wavelength to be specified, we input the measured central wavelengths of each NIRC2 filter into the configuration parameters.  Overall, getting the code running was not too difficult, as there is a handy utility that allows for testing the effect of various input aberrations and comparing the result with the what the model expects, allowing for quick diagnosis and correction of rotation offsets/flip errors in the optics.

\subsection{Off-sky testing and validation}

\begin{figure}[!htb]
\begin{center}
\includegraphics[width=1\textwidth]{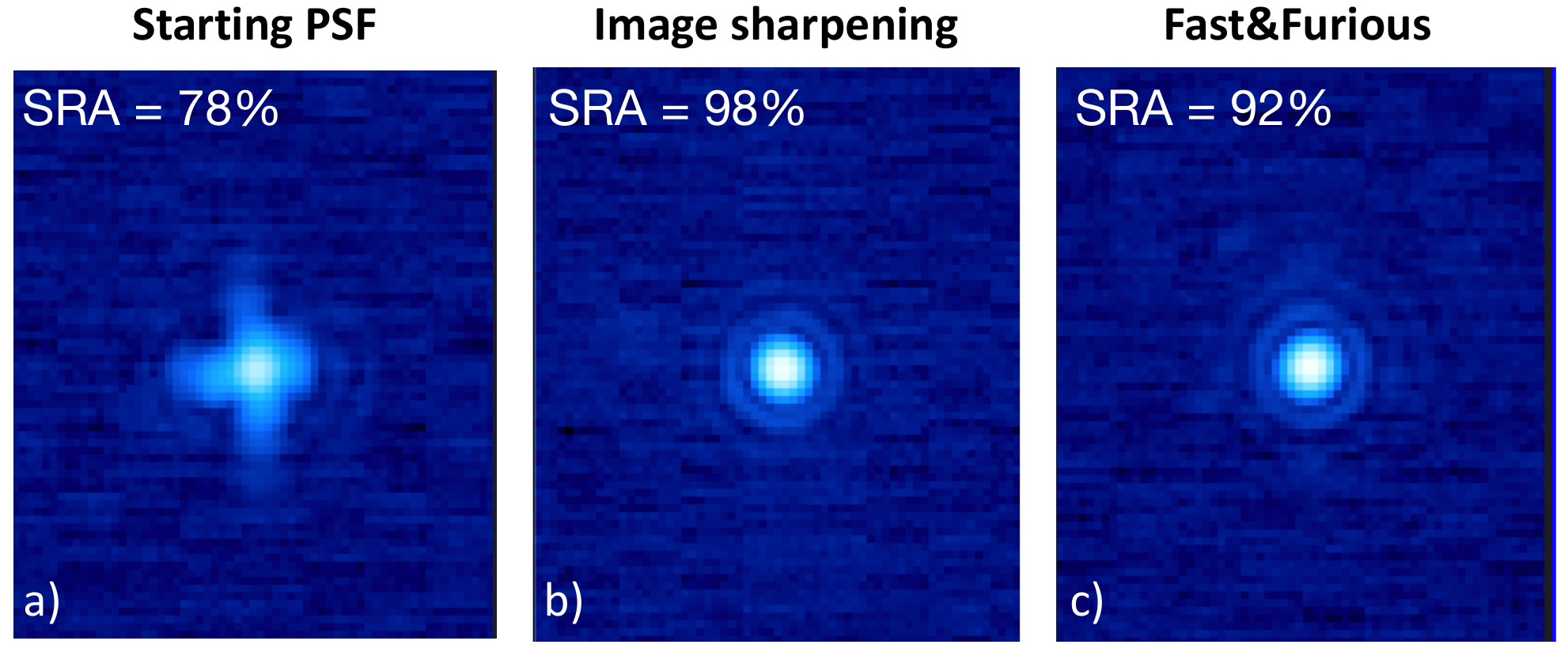}
\end{center}
\caption{
Comparison of the PSFs before and after running the algorithms.
}
\label{fig:PSF_comparison}  
\end{figure} 

\begin{figure}[!htb]
\begin{center}
\includegraphics[width=1\textwidth]{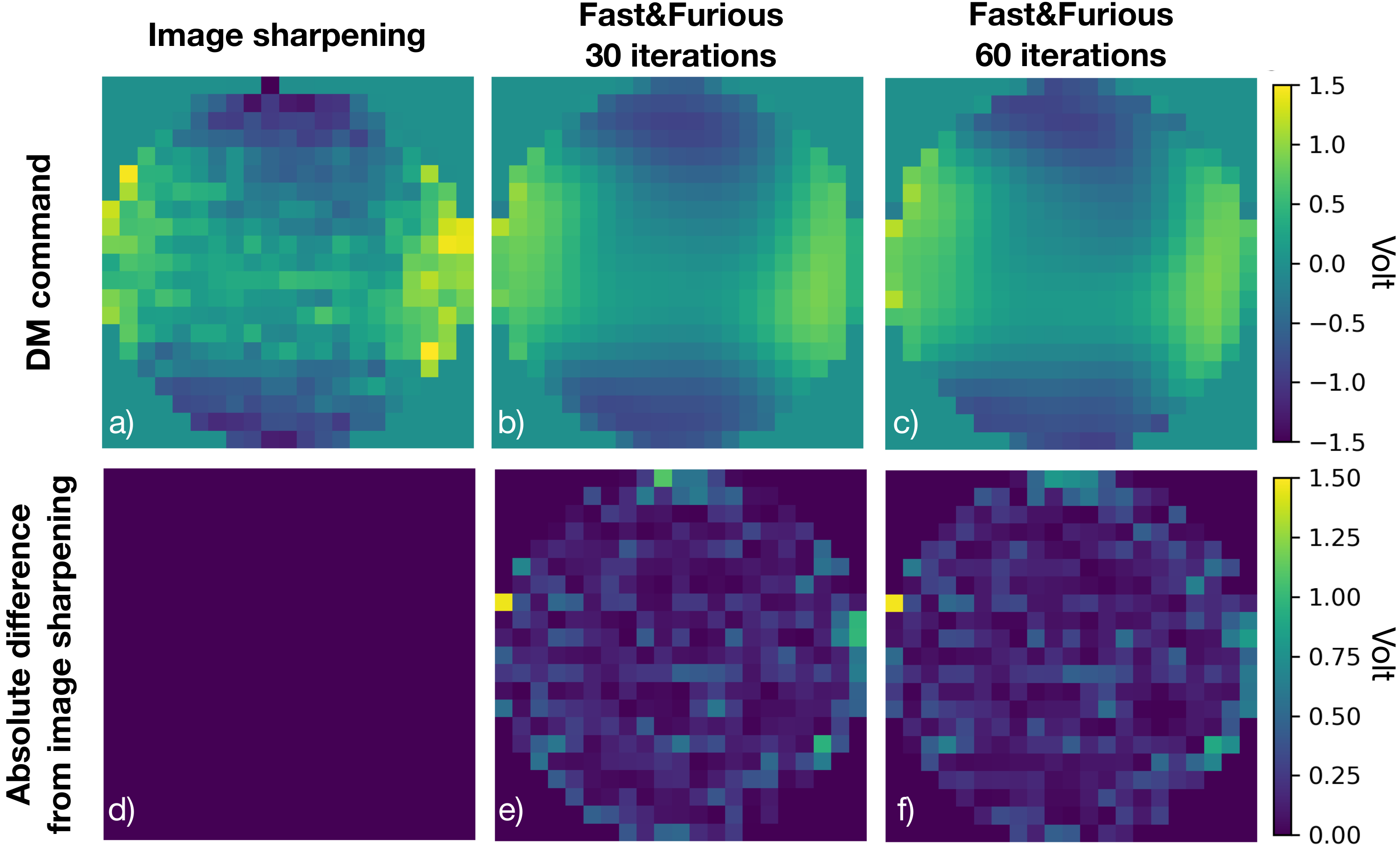}
\end{center}
\caption{
Comparing the DM commands. 
}
\label{fig:DM_comparison}  
\end{figure} 

Once we had installed the code, we ran some initial validation tests comparing the performance against optical simulations.  A similar mismatch in performance was discovered as in Ref~\citenum{bos2020sky} where the convergence rate was at least ten times slower than expected.  Since the slow camera readout speeds required rapid convergence to make this useful, we spent some time trying to understand the issue.  What we eventually discovered was that there was a mismatch in the actual wavefront correction applied by DM compared to what the F\&F algorithm expected, leading to errors in the estimated diversity phase and hence wavefront estimate.  We fixed this issue by multiplying the the DM command by a certain ``boost'' factor to bring it closer to what F\&F was actually expecting, which considerably improved the convergence of F\&F up to the rates demonstrated in simulations.\cite{wilby2018laboratory}  This ``boost'' factor appears to be related to the optical gain of the wavefront sensors, since we found different boost factors between the pyramid and Shack-Hartmann sensor.

With F\&F converging well, we ran a comparison against the in-house Gerchberg-Saxton algorithm.  We input a fixed offset to the deformable mirror that moderately distorted the PSF.  We then ran the standard Gerchberg-Saxton algorithm. Following this, we reloaded the original distortion and ran F\&F as well.  We tried to keep a level of consistency in the experiment.  In both cases, we used the narrow band ($\Delta \lambda = 20.56$ nm) FeII filter at 1.6455 $\mu$m, which is the standard wavelength used for image sharpening.  For F\&F, we set the number of Zernike modes to correct to 90.  All other instrument parameters were identical.

\autoref{fig:PSF_comparison} shows the results of this procedure.  
Both algorithms recovered the same aberration and corrected it.  
The resultant PSFs look similar, though not identical, with Strehl ratios estimated to improve from 78\% to about 98\% for image sharpening and 92\% for F\&F. 
The recovered correction applied to the deformable mirror shown in \autoref{fig:DM_comparison} is consistent,  with noticeable differences only at higher spatial frequencies.  
A statistical look at the actuator voltages revealed that the individual voltages of the ``correction map'' applied to the deformable mirror differed by about 14\% (1-$\sigma$), and these differences were approximately normally distributed. 
Running for 60 iterations did not produce a meaningfully different solution from 30 iterations.
The reason for the different Strehl ratios achieved by the two algorithms is the difference in the number of modes that they correct. 
Image sharpening uses all the degrees of freedom (DoF) of the DM, while F\&F was correcting only 90 low-order Zernike modes, resulting in a much better correction for higher spatial frequencies.
This is also reflected in the difference in applied DM correction as discussed above. 
F\&F can in principle also use all the DM's DoF by not projecting the phase estimate on the Zernike mode basis, but this has not been tested yet on the Keck system. (Strehl ratios above 90\% are rarely achieved on-sky.)

These results are interesting in themselves because the F\&F algorithm is actually easier to implement than the Gerchberg-Saxton from a hardware point of view.  The defocusing in Gerchberg-Saxton requires some method of either moving the source fiber focus, wavefront sensor, or the camera itself, with corrections applied with the deformable mirror.  The F\&F algorithm produces similar results in the same or less time using only the deformable mirror.

\section{On-sky results}\label{sec:results}
We performed on-sky tests in the early morning of December 28$^{\text{th}}$, 2020, using one half night of engineering time.  
Observing conditions were below average by the standards of Maunakea, with seeing at 1.5 - 2 arcseconds, patchy clouds, and high humidity which necessitated multiple closures over our allotted time.
Additionally, the telescope was just two days away from a scheduled segment co-phasing, meaning the level of static aberrations in the system was a bit higher than the norm. 

We initially ran the system with no F\&F loop gain and no introduced aberrations to determine a baseline level from which to compare against.  
In Brackett-gamma, the delivered Strehl ratio (measured from the formula above) from the adaptive optics system was 69.4\% with a 1-sigma variation of 6.5\% over 5 minutes.  
Over the same 30 frames, the Airy ring variation \textit{VAR} was 0.33 with a 1-sigma of 0.1 (in this case, lower is better).  
Unfortunately we did not have the time to thoroughly repeat this measurement at multiple times throughout the night.

In the on-sky testing, we elected to correct 30 Zernike modes, with a system gain of 0.25 (ie, 25\% of the computed correction applied), and a leak factor of 0.95.  We did not use sky background subtraction except in the L' filter.  We also used slightly different wavefront sensor gains between the pyramid and Shack-Hartmann; this is expected due to the different optical gains and sensitivities between the two sensor architectures.

\subsection{Shack-Hartmann wavefront sensor results}
\begin{figure}[!htb]
\begin{center}
\includegraphics[width=1\textwidth]{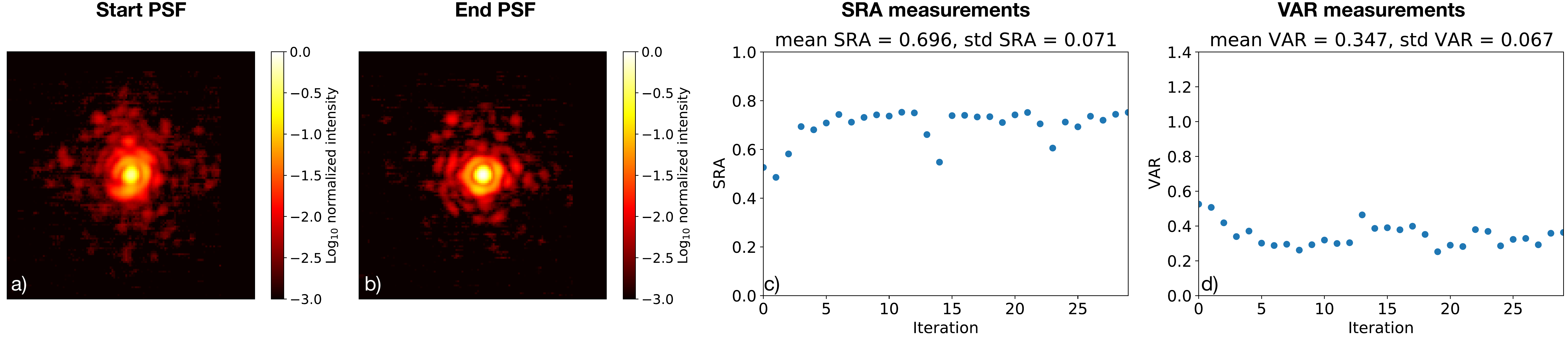}
\end{center}
\caption{
Calibration of the static wavefront error at start of the experiments.
Timestamp: 00:53:41 (HST),
Filter: Bracket-gamma,
Wavefront sensor: Shack-Hartmann
}
\label{fig:static_calibration_SH}  
\end{figure} 
We started the experiment with the Shack-Hartmann operating as the primary wavefront sensor, observing the star HR 2229 (73 Orionis, mag $\sim$5.35 in B to K).
The first test was to measure and correct the static aberrations already present in the system using the Brackett-gamma filter. 
\autoref{fig:static_calibration_SH} shows the results of this test. 
\autoref{fig:static_calibration_SH}a and b show the initial PSF and the PSF after 30 iterations of F\&F operation. 
The PSF noticeably improved after 30 iterations; the first Airy ring is more symmetric and the light is more concentrated at the center. 
These performance resulted in $SRA$ improvement from $\sim$54\% to $\sim$72\%, and $VAR$ improvement from $\sim$0.53 to $\sim$0.33, shown respectively in \autoref{fig:static_calibration_SH} (c) and (d). 
The $SRA$ measurements als show that F\&F quickly converges in 5--10 iterations and then stabilizes. \\
\\
\begin{figure}[!htb]
\begin{center}
\includegraphics[width=1\textwidth]{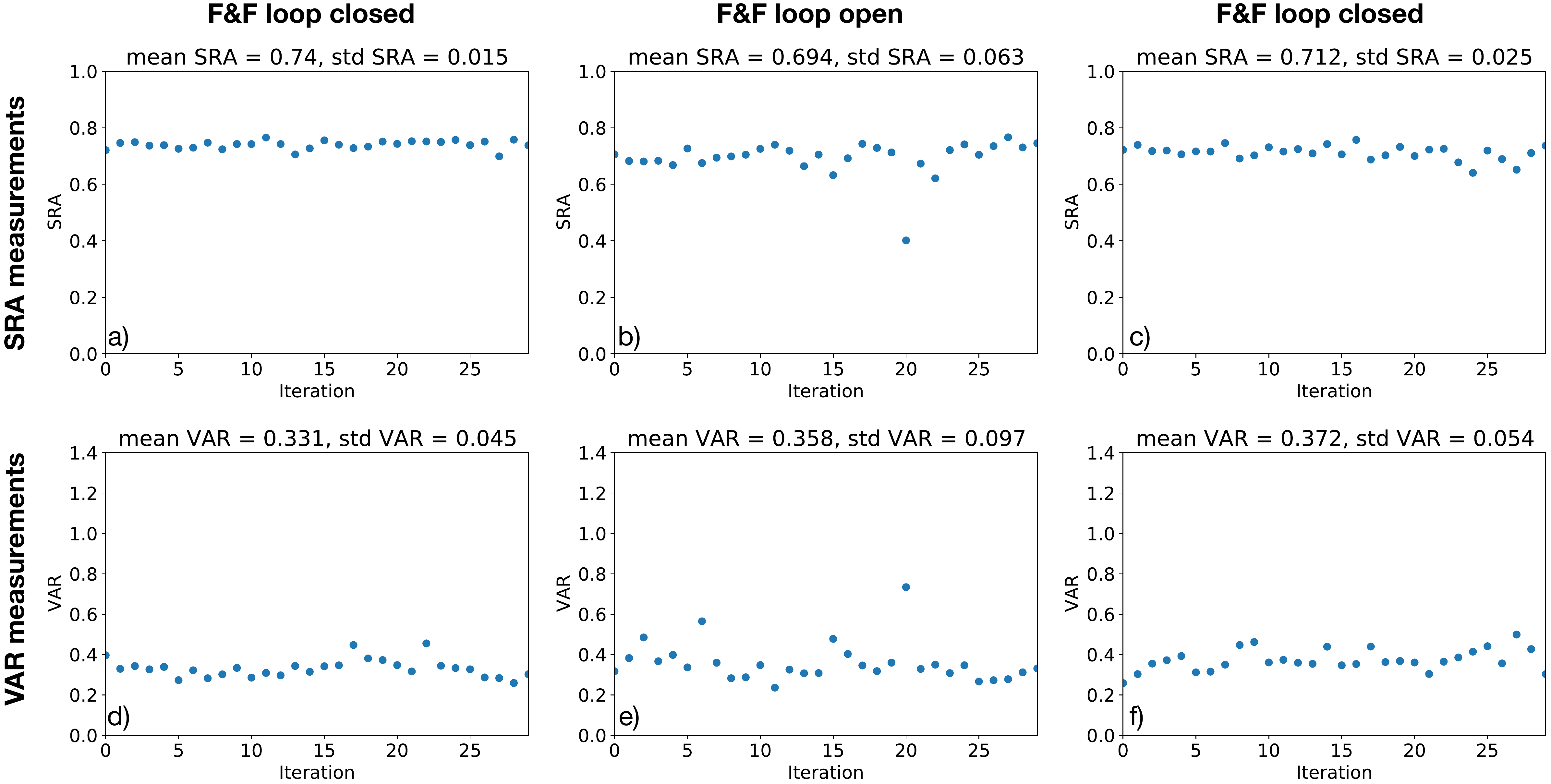}
\end{center}
\caption{
Open- and closed-loop F\&F tests.
Timestamps: 01:07:56, 01:13:11, 01:18:29 (HST), 
Filter: Bracket-gamma,
Wavefront sensor: Shack-Hartmann
}
\label{fig:open_closed_loop_tests}  
\end{figure} 
With the static aberrations corrected, the next test was to investigate to what extend the current F\&F implementation can stabilize the PSF. 
To this end we ran three tests, each of a duration of 30 iterations, where we alternated closed- and open-loop F\&F.   
The $SRA$ and $VAR$ measurements of these tests are shown in \autoref{fig:open_closed_loop_tests}.
During the first closed-loop test the $SRA$ was 74\% with a 1$\sigma$ variation of 2\%, and the $VAR$ was 0.33 a 1$\sigma$ variation of 0.05. 
When the F\&F loop was open, the $SRA$ deteriorated to 69\% with a 1$\sigma$ variation of 6\%, and the $VAR$ worsened to 0.36 a 1$\sigma$ variation of 0.1.
When the F\&F loop closed again, the $SRA$ improved to 71\% with a a 1$\sigma$ variation of 3\%, and the $VAR$ was 0.37 a 1$\sigma$ variation of 0.05. 
These results indicate that when the F\&F loop closes the $SRA$ is higher and more stable, and that the $VAR$ is more stable as well.
However, it is hard to rule out effects from changing atmospheric conditions. \\  
\\
\begin{figure}[!htb]
\begin{center}
\includegraphics[width=1\textwidth]{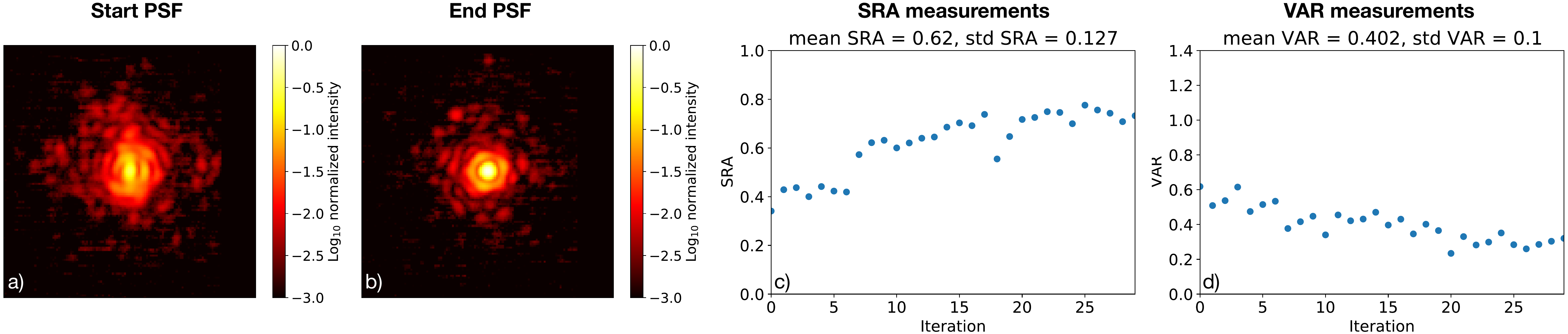}
\end{center}
\caption{
Correction of artificially introduced aberration (with the DM).
Timestamp: 01:26:58 (HST),
Filter: BR gamma,
Wavefront sensor: Shack-Hartmann
}
\label{fig:introduced_aber_BrG}  
\end{figure} 
In \autoref{fig:introduced_aber_BrG} we present a test attempting to correct an artificially introduced aberration consisting of a random combination of low-order Zernike modes projected onto the deformable mirror.
\autoref{fig:introduced_aber_BrG} a and b show the PSF before and after F\&F correction. 
When the aberration is introduced the PSF is strongly distorted, but after 30 iterations of F\&F, the PSF quality has significantly improved. 
\autoref{fig:introduced_aber_BrG}c shows the $SRA$ as a function of iteration.
The introduced aberration lowered the Strehl to 34\%, which improved to 74\% after F\&F correction. 
\autoref{fig:introduced_aber_BrG}d shows the $VAR$ as function of iteration.
The $VAR$ improved from 0.62 to 0.32. \\
\\
\begin{figure}[!htb]
\begin{center}
\includegraphics[width=1\textwidth]{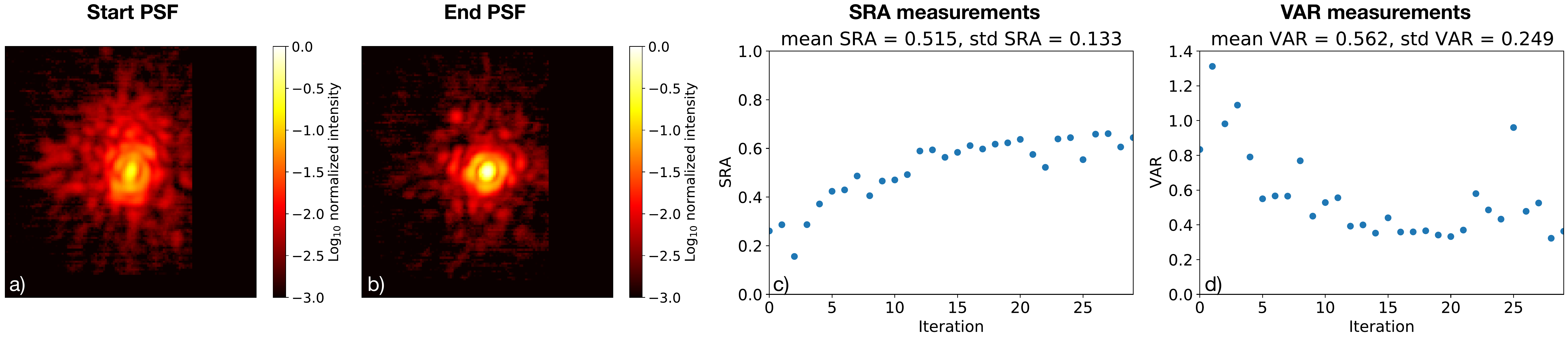}
\end{center}
\caption{
Correction of artificially introduced aberration (with the DM).
Timestamp: 01:40:39 (HST),
Filter: Ks,
Wavefront sensor: Shack-Hartmann
}
\label{fig:introduced_aber_Ks}  
\end{figure} 
To investigate the broadband performance of F\&F we introduced the same artificial aberration with the DM as described above, but in this case changed the filter from Brackett-gamma to K$_\text{s}$. 
The results are shown in \autoref{fig:introduced_aber_Ks}.
\autoref{fig:introduced_aber_Ks}a and b show the PSF before and after correction by F\&F.
The PSF is similarly distorted by the aberration, but more smeared due to the broader filter, which is especially visible at larger separations from the core.
fter the F\&F correction the PSF became more symmetric with more light concentrated in the PSF core. 
\autoref{fig:introduced_aber_Ks}c and show the $SRA$ and $VAR$ as function of iteration, and quantify the extend to which F\&F improves the PSF. 
These results show that F\&F is also capable of operating in broadband filters. \\
\\
\begin{figure}[!htb]
\begin{center}
\includegraphics[width=1\textwidth]{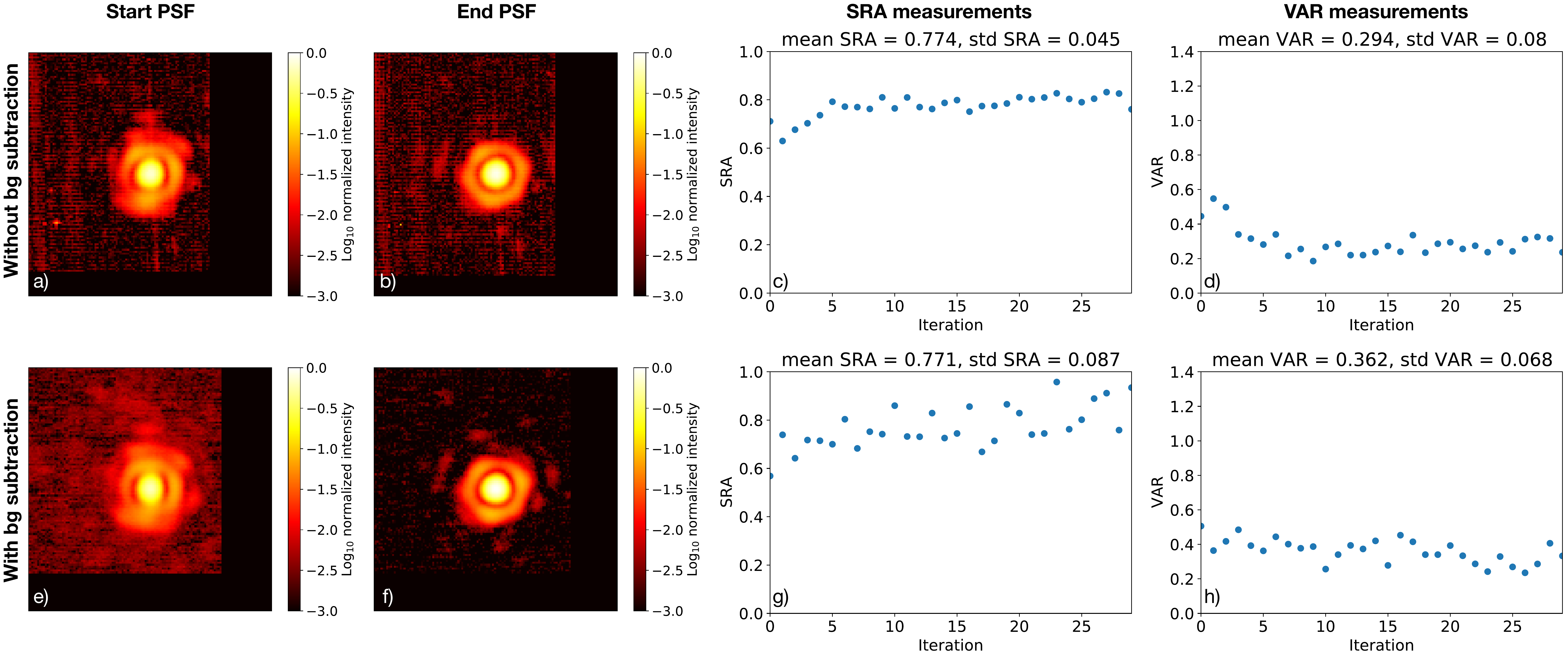}
\end{center}
\caption{
Correction of artificially introduced aberration (with the DM).
Tests without (top row) and with (bottom row) background subtraction. 
Timestamps: 01:46:38, 01:55:26 (HST),
Filter: L',
Wavefront sensor: Shack-Hartmann
}
\label{fig:introduced_aber_Lp}  
\end{figure} 
We next tested F\&F with the L' filter to assess the performance of the algorithm at longer wavelengths, where the thermal background is much higher. 
Here, we tested the algorithm with and without a background subtraction to the image. 
We applied the same static aberrations as in the previous two tests, but due to the longer wavelength of this filter the impact on the PSF was less severe. 
The results of these tests are shown in \autoref{fig:introduced_aber_Lp}. 
The top row shows the results without proper background subtraction and the bottom row with background subtraction. 
The PSFs before and after correction are shown in \autoref{fig:introduced_aber_Lp}a, b, e, f.
For both cases F\&F manages to improve the PSF quality. 
\autoref{fig:introduced_aber_Lp}c, d, g and h show the $SRA$ and $VAR$ measurements during these tests.
These all show that F\&F improves the PSF quality, though it is interesting to note the convergence in the \textit{median} background subtraction test is much better than in the fixed background subtraction test.  This may be due to the background changing quickly enough that subtracting a fixed pattern does not keep up.

\subsection{Pyramid wavefront sensor results}


\begin{figure}[!htb]
\begin{center}
\includegraphics[width=1\textwidth]{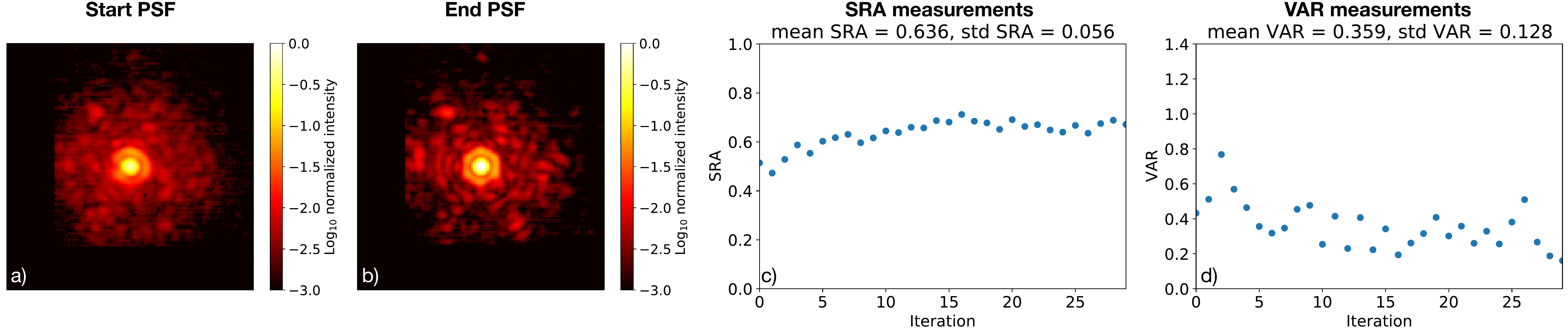}
\end{center}
\caption{
Calibration of the static wavefront error at start of the experiments with the Pyramid WFS.
Timestamp: 04:05:26 (HST),
Filter: Bracket-gamma,
WFS: Pyramid
}
\label{fig:static_calibration_Py}  
\end{figure} 

For the final test, we changed the primary wavefront sensor of the system to the pyramid wavefront sensor, observing the star HR 3086 (85 Geminorum), with magnitudes close (within 0.2 mags) to our previous target.
The different optical paths and gravity vector after the slew resulted into a different static wavefront error compared to the previous tests with the Shack-Hartmann wavefront sensor.
The goal of the test was to correct this static wavefront error with F\&F.
The results are shown in \autoref{fig:static_calibration_Py}.
The PSF before and after F\&F correction is shown in \autoref{fig:static_calibration_Py}a and b, respectively.
After F\&F correction the PSF became more symmetric and the first Airy ring became the hexagonal shape which is expected for the Keck pupil. 
This improvement is quantified by the $SRA$ and $VAR$ measurements presented in \autoref{fig:static_calibration_Py}c an d. 
For both the $SRA$ and the $VAR$ improvements are observed.
Unfortunately, we could not repeat the results of this test; most of the other tests resulted in loop divergence. 
This is likely due to the unfavourable observing conditions and the greater difficulty in applying reference offsets with a pyramid wavefront sensor.
This is supported by the greater difficulty we experienced during internal source tests with F\&F and the pyramid compared to the same tests with the Shack-Hartmann. 
Future work will focus on improving the performance of F\&F with the pyramid wavefront sensor. 
We know from the successful on-sky tests with SCExAO that it is possible to run F\&F with such a wavefront sensor\cite{bos2020sky}. 

\section{Discussion \& Conclusion}\label{sec:conclusion} 

In this work, we have implemented the Fast and Furious algorithm on Keck.  
We first verified its performance against the Gerchberg-Saxton algorithm, finding nearly identical measured aberrations and performance.  
We then tested the algorithm on-sky, with encouraging results, including PSF stabilization and significant Strehl ratio improvements despite poor observing conditions.  
The main differences in this work compared to the previous results in Ref~\citenum{bos2020sky} are the slow loop speeds (dominated by the camera readout time), demonstrations at thermal wavelengths with high sky background, and use of a segmented telescope.  
Despite this, convergence was rapid, with only about 5-15 images needed for significant Strehl ratio improvement, corresponding to only minutes of wall clock time.  
This is encouraging and points to the general utility of the algorithm for AO science, including on telescopes that lack high framerate cameras in their adaptive optics systems.

A few changes are required before F\&F can be integrated into regular on-sky operations at Keck.  
First, the current way to set system parameters is through a text-based interface; a GUI will be needed to make it easier to use.  
Second, the code will need to be integrated with the new real-time control system coming online later this year, which will contain seamless interfaces to both the Shack-Hartmann and pyramid wavefront sensors.
This may go some way to disentangling the confusing results we were getting with the pyramid wavefront sensor.  
Finally, while the current implementation works only in pupil-tracking mode (where the pupil stays fixed and the field rotates), more general use cases may require extending this to field tracking mode.  
This would require real-time tracking of the pupil rotation angle so as to track the rotating aberrations.
This could be a progressive upgrade as the main science cases for F\&F use pupil-tracking mode.

\acknowledgments  
The authors wish to recognize and acknowledge the very significant cultural role and reverence that the summit of Maunakea has always had within the indigenous Hawaiian community.
We are most fortunate to have the opportunity to conduct observations from this mountain.
The research of S.P. Bos leading to these results has received funding from the European Research Council under ERC Starting Grant agreement 678194 (FALCONER).  This material is based upon work supported by the National Science Foundation under Grant No. 2009051 (Advanced Technologies and Instrumentation).  M. Bottom gratefully acknowledges support from the Heising-Simons Foundation. 
This research made use of HCIPy, an open-source object-oriented framework written in Python for performing end-to-end simulations of high-contrast imaging instruments\cite{por2018hcipy}.
This research used the following Python libraries: Scipy\cite{jones2014scipy}, Numpy \cite{walt2011numpy}, and Matplotlib \cite{Hunter:2007}.

\bibliography{report} 
\bibliographystyle{spiebib} 

\end{document}